\documentclass[twocolumn,appendixfloats,tighten]{aastex631}
\usepackage[T1]{fontenc}
\usepackage{ae,aecompl}

\usepackage{amsmath}	
\usepackage{amssymb}	

\usepackage{ctable}
\usepackage{url}
\usepackage{xspace}
\usepackage[normalem]{ulem}
\usepackage{hyperref}
\hypersetup{linkcolor=red,citecolor=blue,filecolor=cyan,urlcolor=magenta}
\usepackage[all]{hypcap}
\usepackage{graphicx}
\usepackage{comment}
\usepackage{color}

\usepackage{amssymb}
\usepackage{pifont}

\usepackage{etoolbox}
\newtoggle{ApJFigs}
\togglefalse{ApJFigs}

\newcommand{\F}{Fig.}

\newcommand{\Eq}{Equation}

\newcommand{\peak}{\mathrm{peak}}
\newcommand{\gw}{\mathrm{GW}}
\newcommand{\orb}{\mathrm{orb}}
\newcommand{\wen}{W03}

\begin{document}

\title{An improved numerical fit to the peak harmonic gravitational wave frequency emitted by an eccentric binary} 

\shortauthors{Hamers}

\author[0000-0003-1004-5635]{Adrian S. Hamers}
\affiliation{Max-Planck-Institut f\"{u}r Astrophysik, Karl-Schwarzschild-Str. 1, 85741 Garching, Germany}

\begin{abstract}
I present a numerical fit to the peak harmonic gravitational wave frequency emitted by an eccentric binary system in the post-Newtonian approximation. This fit significantly improves upon a previous commonly-used fit in population synthesis studies, in particular for eccentricities $\lesssim 0.8$.
\end{abstract}

\section*{}

\begin{figure*}
\iftoggle{ApJFigs}{
\includegraphics[width=0.5\linewidth,trim = 0mm 0mm 0mm 0mm]{figex}
\includegraphics[width=0.5\linewidth,trim = 0mm -3mm 0mm 0mm]{figfit}
\includegraphics[width=0.5\linewidth,trim = 0mm 0mm 0mm 0mm]{figint}
\includegraphics[width=0.5\linewidth,trim = 0mm 0mm 0mm 0mm]{figres}
}{
\includegraphics[width=0.5\linewidth,trim = 0mm 0mm 0mm 0mm]{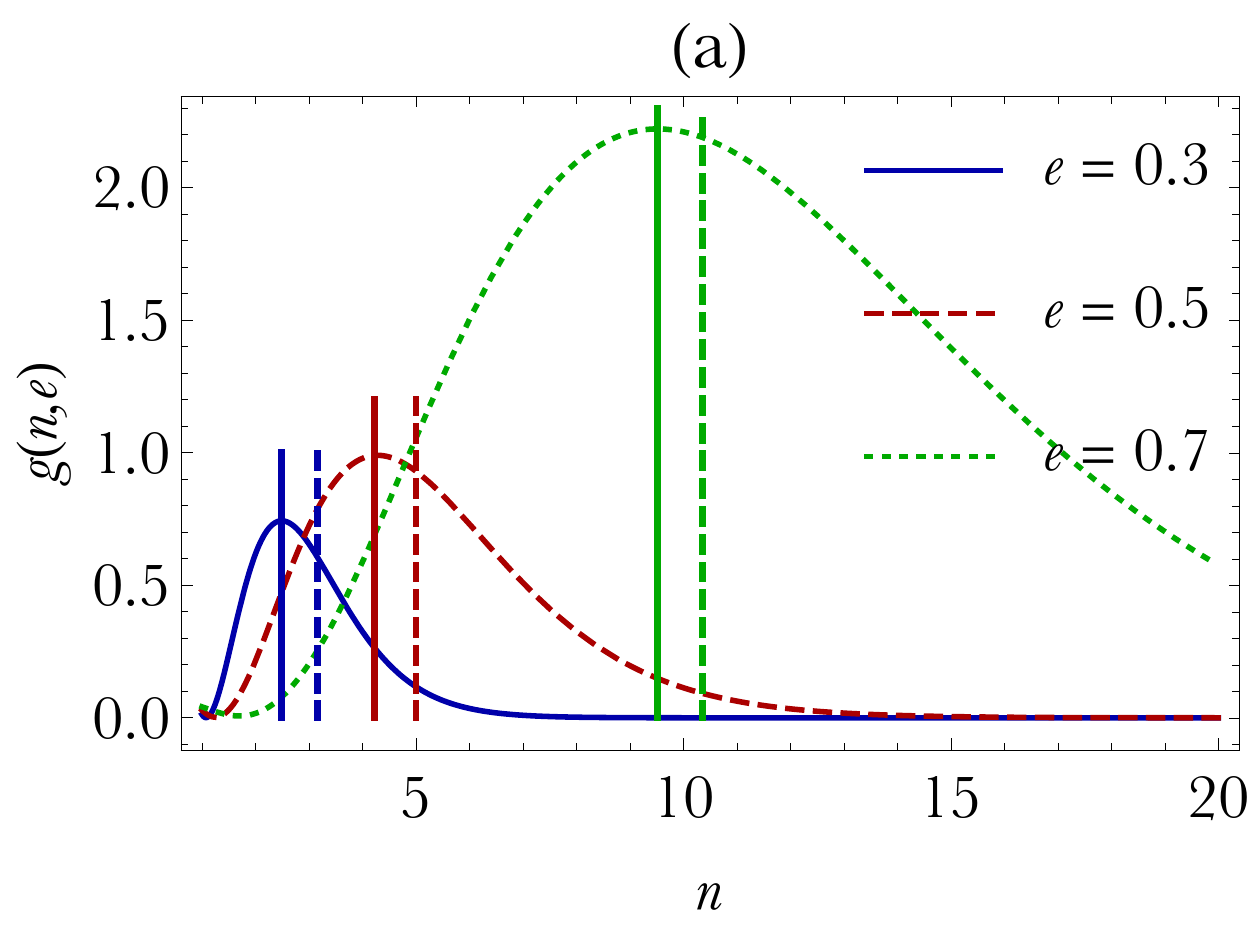}
\includegraphics[width=0.5\linewidth,trim = 0mm -3mm 0mm 0mm]{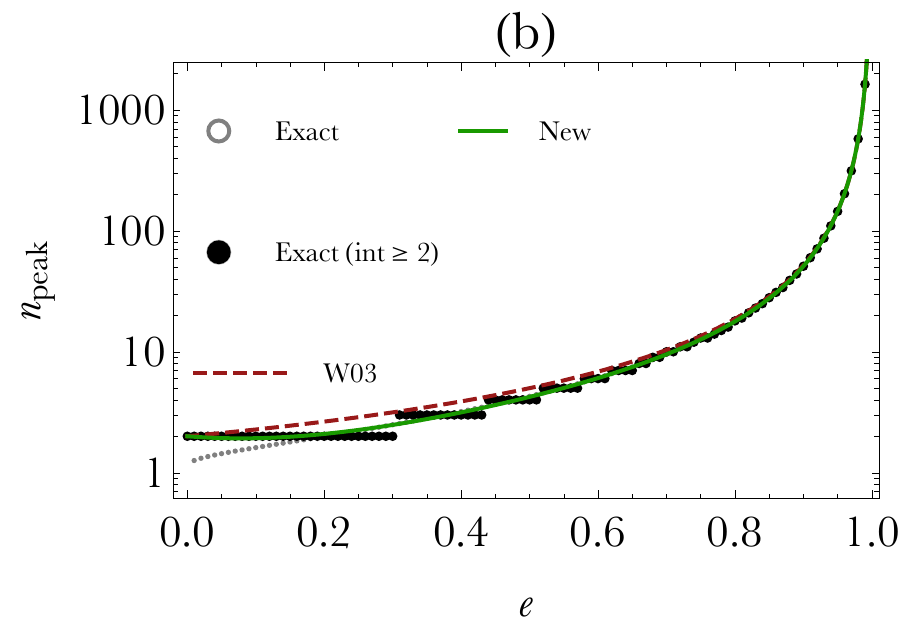}
\includegraphics[width=0.5\linewidth,trim = 0mm 0mm 0mm 0mm]{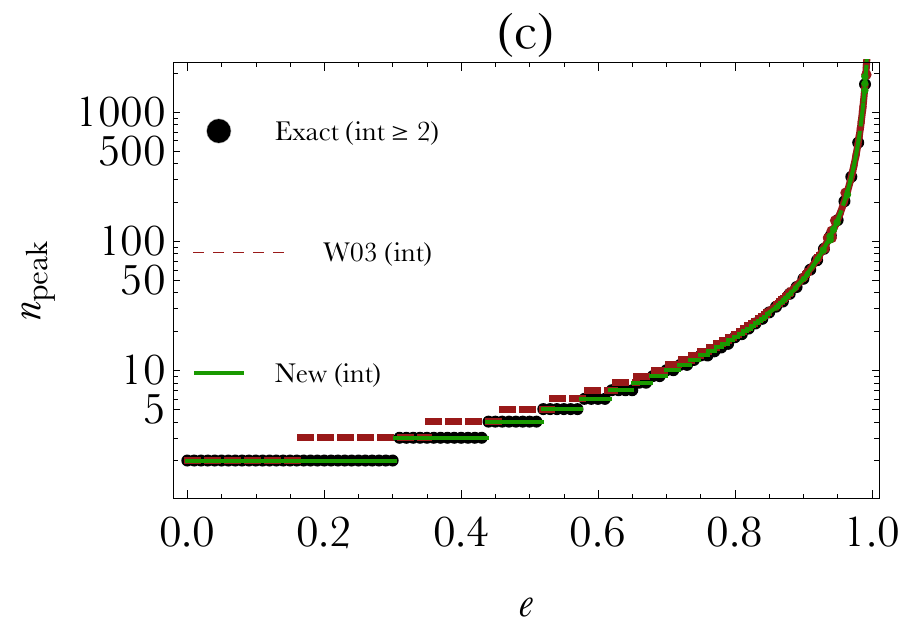}
\includegraphics[width=0.5\linewidth,trim = 0mm 0mm 0mm 0mm]{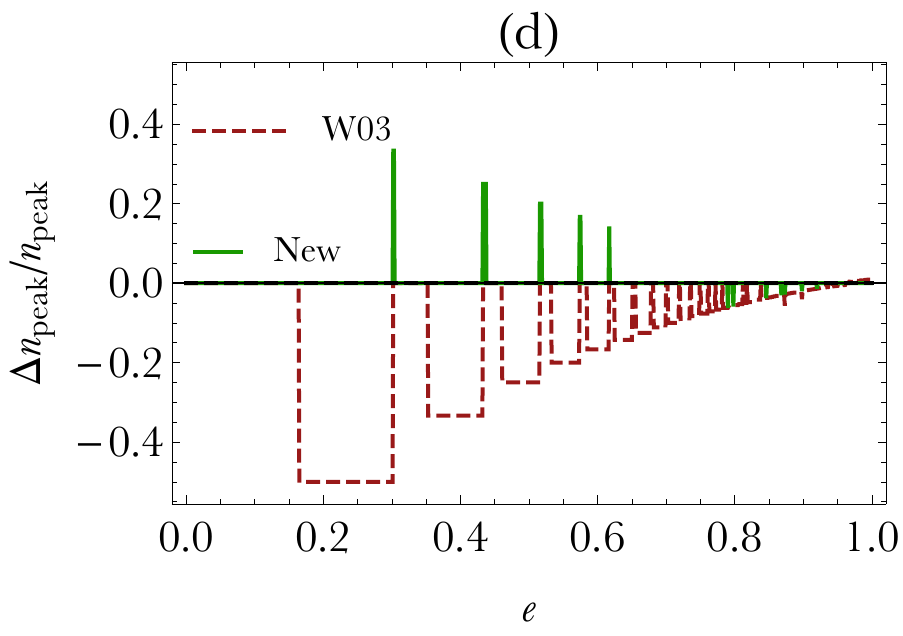}

}
\caption{{\bf(a)} The function $g(n,e)$ (\Eq~\ref{eq:gdef}) as a function of (real) $n$ for several eccentricities. The dashed (solid) vertical lines show the corresponding $n_\peak$ according to the fit of \wen~(\Eq~\ref{eq:nfit}). {\bf(b)} Peak harmonic $n_\peak$ as a function of $e$. Grey open circles show the `exact' calculation of $n_\peak$ by numerically calculating the maximum of $g(n,e)$. The black dots show the exact calculation but with the integer value taken, and imposing $n_\peak \geq 2$. The fit of \wen~is shown with the red dashed line, and the new fit with the green solid line. {\bf (c)} The exact integer values with $n_\peak\geq 2$ are again shown; the two fits are now rounded to integer values. {\bf(d)} Fractional residuals in $n_\peak$ as a function of $e$ (cf. \Eq~\ref{eq:fracdif}). Red dashed line: the fit of \wen; green solid line: the new fit.}
\label{fig}
\end{figure*}

A circular binary system (semimajor axis $a$, masses $m_1$ and $m_2$) emits gravitational waves (GWs) at a frequency $f_\gw$ equal to twice the orbital frequency $f_\orb$, i.e., $f_\gw = 2f_\orb$, where 
\begin{align}
f_\orb = \frac{1}{2\pi} \sqrt{\frac{GM}{a^3}}.
\end{align}
Here, $M\equiv m_1+m_2$ is the total binary mass, and $G$ is the gravitational constant. As shown in the seminal work of \citet{1963PhRv..131..435P} that assumed the (lowest-order) post-Newtonian approximation, an eccentric binary (eccentricity $e$) emits GWs at additional harmonics. Specifically, the power of the $n^\mathrm{th}$ harmonic with frequency $f_{\gw,n} = n f_\orb$ (with integer $n \geq 1$) is given by
\begin{align}
P_n = \frac{32}{5} \frac{G^4}{c^5} \frac{m_1^2 m_2^2 M}{a^5} g(n,e),
\end{align}
where the function $g(n,e)$ quantifies the factor to which more power is emitted in the $n^\mathrm{th}$ harmonic compared to a circular orbit. The latter function is given by
\begin{align}
\label{eq:gdef}
\nonumber &g(n,e) = \frac{n^4}{32} \biggl \{ \biggl [ J_{n-2}(ne) - 2e J_{n-1}(ne) + \frac{2}{n} J_n(ne) \\
\nonumber &\quad + 2e J_{n+1}(ne) - J_{n+2}(ne) \biggl ]^2 + \left(1-e^2\right) \biggl [J_{n-2}(ne) \\
&\quad- 2 J_n (ne)+J_{n+2}(ne) \biggl ]^2 + \frac{4}{3n^2} J_n(ne)^2 \biggl \},
\end{align}
where $J_i(x)$ is the $i^\mathrm{th}$ Bessel function of the first kind. \F~\ref{fig}(a) plots $g(n,e)$ as a function of $n$ (interpreted as a real number) for several values of $e$. More eccentric binaries typically emit more power at higher harmonics. 

It is of interest to consider the peak harmonic, $n_\peak$, i.e., $n_\peak(e)$ is the value of $n$ for which $g(n,e)$ is at a maximum. The corresponding peak GW frequency immediately follows from the relation
\begin{align}
f_{\gw,\peak}(e) = n_\peak(e) f_\orb.
\end{align}
It is generally cumbersome to numerically determine $n_\peak(e)$ from \Eq~(\ref{eq:gdef}). \citeauthor{2003ApJ...598..419W} (\citeyear{2003ApJ...598..419W}, hereafter \wen) provided a convenient fitting function to $n_\peak(e)$ given by
\begin{align}
\label{eq:nw03}
n^{\mathrm{(W03)}}_\peak(e) = 2 \frac{(1+e)^{1.1954}}{\left(1-e^2\right )^{3/2}}.
\end{align}
\Eq~(\ref{eq:nw03}) is plotted as dashed vertical lines in \F~\ref{fig}(a). Although a reasonable fit for high eccentricities, it does not capture the true maximum of $g(n,e)$ very accurately for lower eccentricities. Nevertheless, the fit of \wen~is commonly used to estimate the peak GW frequency of eccentric binaries, in particular in the context of population synthesis studies \citep[e.g.,][]{2011ApJ...741...82T,2012ApJ...757...27A,2014ApJ...784...71S,2018PhRvD..98l3005R,2018ApJ...865....2H,2021ApJ...917...28K,2021ApJ...920...81S,2021arXiv211014680V}. 

Here, I present a new numerical fit of $n_\peak(e)$ which significantly improves upon that of \wen~for low eccentricities ($e \lesssim 0.8$), whereas also accurate for high eccentricities:
\begin{align}
\label{eq:nfit}
n_\peak(e) \simeq 2\left (1 + \sum_{k=1}^4 c_k e^k \right )\left(1-e^2 \right )^{-3/2},
\end{align}
where $c_1=-1.01678$, $c_2 = 5.57372$, $c_3 = -4.9271$, and $c_4 = 1.68506$. The fit retains the factor $(1-e^2)^{-3/2}$ from \wen~that dominates at high eccentricities, but the behavior at smaller eccentricities is modified. \Eq~(\ref{eq:nfit}) correctly states that $n_\peak(0)=2$ for circular orbits.

The new fits are indicated in \F~\ref{fig}(a) with solid vertical lines, and show significantly better match with the true $n$ corresponding to a maximum in $g(n,e)$ (when $n$ is interpreted as a real number). 

\F~\ref{fig}(b) plots the peak harmonic $n_\peak$ as a function of $e$. Grey open circles show the `exact' calculation of $n_\peak$ by numerically calculating the maximum of $g(n,e)$ for given eccentricity. The latter is carried out in practice for real $n\geq 1$, although it should be understood that $n$ is actually an integer. The exact (real) calculation yields that $n_\peak$ decreases below 2 as $e\rightarrow 0$. However, it is clear that, in the circular limit, $g(n,0)=1$ for $n=2$, and $g(n,0)=0$ for all other integer $n$. Therefore, when determining $n_\peak$, one should take the integer value and limit to $n_\peak\geq 2$ in order to retain the correct behavior in the fitting function as $e\rightarrow 0$.

By enforcing that $n_\peak(0)=2$ and given the limitations of the assumed functional form, the fit of \wen~significantly overpredicts $n_\peak$ for $e \lesssim 0.8$. The new fit better captures the low-eccentricity regime, while also still satisfying $n_\peak(0)=2$ and giving a good description at high eccentricities. This is particularly clear in \F~\ref{fig}(c), in which rounded integer numbers are plotted. 

Lastly, \F~\ref{fig}(d) shows the fractional residuals in the integer $n_\peak$ computed as the difference between the `exact' integer-rounded calculation with $n_\peak \geq 2$ and the integer-rounded fits, i.e.,
\begin{align}
\label{eq:fracdif}
\frac{\Delta n_\peak}{n_\peak} = \frac{ \mathrm{int} \left [ n^{(\mathrm{exact})}_\peak \right ] - \mathrm{int} \left [ n^{(\mathrm{fit})}_\peak \right ]}{\mathrm{int} \left [ n^{(\mathrm{exact})}_\peak \right ] }.
\end{align}
For $e \lesssim 0.8$, the fit of \wen~systematically overpredicts $n_\peak$ ($\Delta n_\peak < 0$). This is especially the case for the range $0.17 \lesssim e \lesssim 0.30$, where \Eq~(\ref{eq:nw03}) predicts $n_\peak=3$, whereas it should be $n_\peak=2$. For larger $e$, the discrepancies become less severe, although at $e=0.8$ the peak harmonic is still overpredicted by $\sim 10\%$. For even higher $e$, $e\gtrsim 0.97$, \Eq~(\ref{eq:nw03}) starts to slightly underpredict $n_\peak$ ($\Delta n_\peak > 0$). 

In contrast, the new fit \Eq~(\ref{eq:nfit}) has typically zero fractional residuals, with only a few spikes occurring due to rounding effects at transitions where $n_\peak$ advances by unity. At $e=0.999$ (the highest eccentricity considered when determining \Eq~\ref{eq:nfit}), the new fit has zero fractional residuals, i.e., $n_\peak(0.999)=51,755$ according to \Eq~(\ref{eq:nfit}) is consistent with the exact calculation. In contrast, \Eq~(\ref{eq:nw03}) has a fractional error of $\simeq 1\%$ with $n^{(\mathrm{W03})}_\peak(0.999)=51,216$.

When applying the fit presented here, it should be remembered that it relies on the results of \citet{1963PhRv..131..435P} based on the lowest-order post-Newtonian terms that describe dissipation due to GW emission (2.5PN). Higher-order post-Newtonian corrections \citep[e.g.,][]{2021PhRvD.104j4023T} are not included.

Also to be considered in \Eq~(\ref{eq:nfit}) is that the peak GW frequency does not describe the amplitude, or, more relevantly, the signal-to-noise ratio in the GW detector band. The latter should also be taken into account when making statements about detectability of GW sources (using, e.g., LEGWORK, \citealt{2021arXiv211108717W}). 

Lastly, the fits considered by \wen~and the new fit here quantify the energy flux, but not the angular momentum flux, which could peak at a different harmonic. The latter should be considered in future work, as it controls orbital circularisation.

\begin{acknowledgements} 
I thank Selma de Mink, Tom Wagg, Alejandro Vigna-G\'{o}mez, and an anonymous referee for helpful comments. 
\end{acknowledgements}

\bibliographystyle{aasjournal}
\bibliography{literature.bib}

\end{document}